\documentclass[12pt]{article}

\begin{document}
\begin{titlepage}
\title{Quantum spherical spin-glass with random short-range
interactions}
\author{Pedro Castro Menezes\\
        Alba Theumann\thanks{albath@if.ufrgs.br}\\
        Instituto de Fisica\\
        Universidade Federal do Rio Grande do Sul\\
        Avenida Bento Gon\c{c}alves 9500, C.P. 15051\\
        91501-970, Porto Alegre, RS, Brazil}
\date{}
\maketitle
\thispagestyle{empty}
\end{titlepage}
\begin{abstract}
\normalsize \noindent  In the present paper we analyze the
critical properties of a quantum spherical spin glass model with
short range, random interactions. Since the model allows for
rigorous detailed calculations, we can show how the effective
partition function calculated with help of the replica method for
the spin glass fluctuating fields $ Q_{\alpha \gamma}(\vec{k}
\omega_1 \omega_2) $ separates into a mean field contribution for
the $Q_{\alpha \alpha}(0; \omega ; -\omega) $ and a strictly short
range partition function for the fields $ Q_{\alpha \neq
\gamma}(\vec{k} \omega_1 \omega_2) $.Here $ \alpha, \gamma =1..n $
are replica indices.  The mean field part $ W_{MF}$ coincides with
previous results. The short range part $W_{SR}$ describes a phase
transition in a $ Q^3 $-field theory, where the fluctuating fields
depend on a space variable $\vec{r} $ and two times $ \tau_1$ and
$\tau_2$. This we analyze using the renormalization group with
dimensional regularization and minimal subtraction of dimensional
poles. By generalizing standard field theory methods to our
particular situation, we can identify the critical dimensionality
as $ d_c = 5$ at very low temperatures due to the dimensionality
shift $D_c = d_c+1 =6 $. We then perform a $ \epsilon^\prime $
expansion to order one loop to calculate the critical exponents by
solving the renormalization group equations.
\end{abstract}
\newpage
\section*{1.\ Introduction}
Since the formulation of the renormalization group theory to explain
the critical behaviour and scaling properties of phase transitions,
it emerged the natural question of how this theory would apply to
phase transitions in quantum systems. In these systems, time plays
an essential role through the equations of motion of the operators
even in equilibrium quantum statistical mechanics, then a natural
conjecture was that there would be a dimensional shift from the
space dimension $d$ to the effective $D= d+1$, and that scaling
behaviour in the critical region would require the introduction of a
new critical dynamical exponent $z$ \cite{zazie,hertz}. This is
evident when the quantum mechanical partition function is written as
a functional integral in terms of fields that are functions of
position and imaginary time $\tau $ variables, where $ 0 \le \tau
\le \beta $ and $\beta= \frac{1}{T} $ is the inverse temperature in
units with $ \hbar=k_B=1 $, as phase transitions occur in infinite
systems when the correlation length $\xi $ becomes infinite at the
critical temperature $T_c$. If $T_c > 0$, the "length " $\beta_c $
in the imaginary time direction is finite and the associated
correlation length $ \xi^{z} > \beta_c $, then the transition would
be classical in $d$ space dimensions. However, if quantum
fluctuations drive the critical temperature to $ T_c =0$, at this
point the time length $\beta_c$ is infinite and a new transition
with a dimensional shift $D=d+1 $ is expected at a quantum critical
point (QCP)
\cite{belitz}.\\
Physical realizations of quantum phase transitions occur in strongly
correlated systems\cite{mucio} and other physical systems as
described extensively in ref.(\cite{sachdev}). A particular class of
systems that present a quantum critical point are quantum spin
glasses like the insulating $ LiHo_{x}Y_{1-x}F_{4}$
\cite{experiments}, that is well represented by the $M$-component
spin-glass model in a transverse field\cite{edu} or by the
spin-glass model of $ M$-components quantum
rotors\cite{sachdev,readsachdev}. In the limit $M \rightarrow \infty
$ the quantum rotors  model\cite{ye} reduces to the quantum
spherical model for a spin glass that has been studied before
\cite{shukla} in the mean field limit of infinite range
interactions, following the classical spin glass theory of
Sherrington-Kirkpatrick (SK)\cite{SK}. It was also shown that the
effective action of the quantum spherical spin glass is
invariant\cite{pedroalba} under the Becchi-Rouet-Stora-Tyutin
supersymmetry and consequently the spin glass order parameter
vanishes while replica symmetry (RS) is exact. The $p$-spin quantum
spherical model was also studied by using the boson operators
representation for the harmonic oscillator\cite{theo,serral} and it
was shown that the systems with $p=2$  and $p\ge 3$ belong to
different universality classes. For $p\ge 3$ there is replica
symmetry breaking (RSB) and the system belongs to the same
universality class as the SK model, with a finite order parameter $Q
\neq 0$ below the critical
temperature.\\
Since the formulation of (SK) spin-glass theory with infinite
range interactions, the natural question was asked of how finite
range, random interactions would modify the critical properties of
spin glasses and which would be the critical exponent associated
to them. To answer this question, renormalization group
calculations were performed above criticality \cite{harris} in an
expansion in $ \epsilon = 6-d $ for short range interactions, from
where emerges $d_c = 6$ as the critical dimensionality of the
classical spin-glass. Renormalization group calculations for long
range interactions decaying as $ r^{-(d+\sigma)} $ in the
classical spin-glass were also performed\cite{kotliar,sak}. Below
the critical temperature there is replica symmetry breaking(RSB)
and a non-vanishing order parameter that is in fact a matrix in
replica space, then a more difficult renormalization group in
replica space should be performed, as it is discussed in detail in
ref.(\cite{cdd}).Here also the critical dimensionality appears to
be $d_c = 6$, thus completing the description of the short-range
classical spin glass  below the critical temperature.\\
 It is
the purpose of this paper to analyze the critical properties of a
quantum spherical spin glass with random short range interactions
by using renormalized perturbation theory with dimensional
regularization and minimal subtraction of dimensional
poles\cite{amit}. This task is far from trivial as the structure
of the resulting field theory differs from standard theories. To
start with, the quantum rotor model without disorder is covariant
in space and time\cite{sachdev}, the frequency appearing just as a
new momentum component to form a (d+1)-dimensional vector with
modulus $ k^2 + \omega^2 $, giving thus the value of the exponent
$ z=1 $, but this is not the case in the disordered model. The
disorder has no dynamical fluctuations and if average over
disorder restores space translational invariance, this is not
necessarily so in the time direction. Consequently, we are forced
to consider an effective action in terms of a spin glass field
that depends on position $\vec{r} $ and on two imaginary times
$\tau_1$ and $\tau_2$ (not on the two times difference). This
carries on the need to formulate in this paper new rules for the
calculation of diagrams in the loop expansion\cite{amit}. We find
accordingly that under renormalization the relation between space
and imaginary time changes and as a result the exponent $z$
differs
from unity.\\
The quantum spherical model for a spin glass is the ideal testing
ground for these ideas, as its formulation in terms of functional
integrals allows for rigorous analytic calculations. As a difference
with the infinite range quantum spherical spin
glass\cite{pedroalba}, in the case of short range interactions we
have to use the replica method to derive the effective action, and
this we do in Sect.2. In Sect.3 we derive the renormalized
perturbation theory to order one loop in an expansion in $
\epsilon^\prime = D-6= d-5 $, then our critical space dimension is
effectively $d_c = 5$. We calculate the critical exponents by
solving the renormalization group equations in the critical region.
We leave Sect.4 for discussions. The detailed and far from trivial
calculations are described in the Appendix to keep the natural flow
of calculations in the paper.
\newpage
\section*{2.\ The model}
We consider a spin glass of quantum rotors with moment of inertia
$I$ in the spherical limit\cite{ye,pedroalba,shukla}with Hamiltonian
\begin{eqnarray}
{\cal H}_{SG}+\mu \displaystyle \sum_{i}S_{i}^2 = \frac{1}{2I}
\displaystyle \sum_{i} P_{i}^2 - \displaystyle \frac{1}{2}
\sum_{i,j}J_{ij}S_{i}S_{j} + \mu \displaystyle \sum_{i}S_{i}^2
\label{1}
\end{eqnarray}
\noindent where the spin variables at each site are continuous $
-\infty < S_{i} < \infty $ and we introduced
the canonical momentum $ P_{i}$ with commutation rules:\\
\begin{eqnarray}
[S_{j},P_{k}] = i \delta_{j,k} \label{2}
\end{eqnarray}
\noindent The sum  in eq.(\ref{1})runs over sites $ i,j =1..N $.
The bond coupling $J_{ij}$ in eq.(\ref{1}) is an independent
random variable with the gaussian
distribution\cite{kotliar,harris}
\begin{eqnarray}
P(J_{ij})=e^{-\frac{J_{ij}^{2}}{2J^{2}V_{ij}}}\sqrt{\frac{1}{2\pi
J^{2}V_{ij}}} \label{3}
\end{eqnarray}
and $ V_{ij} = V(\vec{R_i} -\vec{R_j})$ is a short range
interaction with Fourier transform at low momentum $ k $
\begin{eqnarray}
V(k) \approx 1-k^2 \label{4}
\end{eqnarray}
 The chemical potential $\mu$ is a Lagrange multiplier that
insures the mean spherical condition
\begin{eqnarray}
-\frac{\partial\langle ln{\cal W}\rangle}{\partial(\mu)}=
\displaystyle \sum_{i}\int_0^{\beta} d\tau \langle S_{i}^2 \rangle =
\beta N \label{5}
\end{eqnarray}
and $ \beta = 1/T $ is the inverse temperature. We work in units
where the Boltzmann constant $ k_B = \hbar= 1$ and ${\cal W}$ is
the quantum partition function
\begin{eqnarray}
&&{\cal W}= Tre^{-\beta({\cal H}_{SG}+\mu \sum_{i}S_{i}^2 )}
\label{6}
\end{eqnarray}
 That can be expressed as a functional
integral\cite{pedroalba,negele,feynman}\\
\begin{eqnarray}
{\cal W}= &&\displaystyle \int \prod_{i}{\cal D} S_{i}
\exp{\left(-\cal{ A_{O} - A_{SG}} \right)} \label{7}
\end{eqnarray}
where the non interacting action $ \cal A_{O} $ is given by
\begin{eqnarray}
{ \cal A_{O}} =\int_{0}^{\beta} d \tau  \displaystyle
\sum_{i}\left(\frac{I}{2} \left(\frac{\partial S_{i}}{\partial
\tau}\right)^{2} + \mu S_{i}^2(\tau)\right) \label{8}
\end{eqnarray}
and the interacting part
\begin{eqnarray}
{ \cal A_{SG}} =\frac{1}{2}
\displaystyle\sum_{i,j}J_{ij}\int_{0}^{\beta} d \tau
S_{i}(\tau)S_{j}(\tau) \label{9}
\end{eqnarray}
 The free energy may be calculated with the replica method
\begin{equation}
F=-\frac{1}{\beta N}\lim_{n\rightarrow 0}\frac{W_{n}-1}{n}
\label{10}
\end{equation}
where  $< {\cal W}^n>_{ca}=W_n$ is the partition functional for
$n$-identical replicas, configurationally averaged over the
probability distribution of $J_{ij}$ in eq.(\ref{3}). It is shown
in the Appendix that $W_n$ may be expressed as a functional over
fluctuating spin glass fields $ Q_{\alpha
\gamma}(\vec{k},\omega,\omega^\prime) $, where $\omega=\frac{2\pi
m}{\beta} $ is a discrete Matsubara frequency for finite
temperature and $\alpha,\gamma=1..n $ are replica indices. The
result obtained in eq.(\ref{46}) is that the partition functional
separates into two parts
\begin{equation}
W_n = W_{MF} W_{SR} \label{11}
\end{equation}
where $ W_{MF} $ in eq.(\ref{47}) is the mean field functional for
the fields $Q_{\alpha \alpha}(0,\omega,-\omega) $ already obtained
in ref.(\cite{pedroalba}) that determines the critical temperature
$ T_c(I) $, while $Z_{SR}$ depends on the spin glass fluctuations
$ Q_{\alpha \neq \gamma}(\vec{k},\omega,\omega^\prime) $ for short
range interactions and determines the critical behaviour. We
remark that these fields depend naturally on two independent times
(frequencies) and not on the difference of two times, because the
disorder is not time correlated and it restores translational
invariance in space, but not in time\cite{readsachdev}.We obtain
from eq.(\ref{49})
\begin{equation}
 W_{SR}= \int\displaystyle \prod_{\alpha \neq \gamma}{\cal{D}}
 Q_{\alpha \gamma}(\vec{k},\omega,\omega^\prime)
 \exp{(-A_{SR}\{Q\})}
 \label{12}
\end{equation}
where $\alpha ,\gamma =1..n $ are replica indices and
\begin{eqnarray}
A_{SR}\{Q\}=
 \displaystyle \sum_{\alpha \neq
\gamma}\sum_{\omega_1 \omega_2}\int d\vec{k}
\Big[\frac{\mu-\mu_c}{\mu_c} +k^2+
s^2(\omega_1^2+\omega_2^2)\Big]Q_{\alpha
\gamma}(\vec{k},\omega_1,\omega_2)Q_{\alpha
\gamma}(-\vec{k},-\omega_1,-\omega_2) \nonumber \\
+ \frac{\lambda}{3!}\displaystyle \sum_{\alpha \neq \gamma \neq
\delta}\sum_{\omega_1 \omega_2 \omega_3}\int d\vec{k_1}d\vec{k_2}
Q_{\alpha \gamma}(\vec{k_1},\omega_1,\omega_2)Q_{\gamma \delta
}(\vec{k_2},-\omega_2,\omega_3)Q_{ \delta
\alpha}(-\vec{k_1}-\vec{k_2},-\omega_3,-\omega_1)\nonumber\\
\label{13}
\end{eqnarray}
Having in mind a renormalization group calculation, the frequency
term in the non-interacting inverse propagator is affected by the
coefficient $s$, as it will turn out that momentum and frequency
renormalize differently and they cannot be kept both equal to
unity.The infinite volume limit was taken in eq.(\ref{13}), but for
the moment the temperature is kept finite and the sums are over
discrete Matsubara frequencies. In all the following work is
implicit that $ Q_{\alpha \gamma} $ means $  Q_{\alpha \neq
\gamma}$, while $Q_{\alpha}(\omega) $ means $  Q_{\alpha
\alpha}(0,\omega,-\omega)$.\\
We now proceed with the renormalization group calculation using
dimensional regularization and minimal subtraction of dimensional
poles\cite{amit}, to one loop order. In eq.(\ref{13}) we kept only
the terms $ O(Q^3)$ because the terms $ O(Q^4)$ would be irrelevant
close to the critical dimensionality of a $ Q^3$-theory, as there is
no change in the sign of $\lambda$ for the gaussian probability
distribution of the random bonds\cite{harris}. To analyze the value
of the critical dimensionality we consider separately the case of
finite temperature than that of $ T=0$. In both cases the vertex
functions that present divergencies needing renormalization are the
inverse propagator $\Gamma^{(2)}$, the three point vertex function
$\Gamma^{(3)}$ and the two point vertex function with one insertion
$\Gamma^{(2,1)}$\cite{amit}. To one loop order they are given by the
diagrams in fig.(2). At this point it is important to distinguish
between the system temperature $T$ and the critical parameter $
t=\frac{\mu -
\mu_c}{\mu_c} $ that measures the approach to criticality.\\
We start by analyzing the transition at finite temperature $T$. The
action in eq.(\ref{13}) must be dimensionless, then dimensional
analysis tells us that, for $\Lambda$ an inverse lenght
\begin{eqnarray}
[k]=\Lambda\hspace{0.5cm}[Q]=\Lambda^{-d/2-1}\hspace{0.5cm}[\lambda]=\Lambda^{3-d/2}
\label{14}
\end{eqnarray}
and the critical dimensionality is $ d_c=6$, as corresponds to a
classical system. The vertex functions calculated with the usual
rules in $\varphi^3$-field theory \cite{amit,theumanngusmao} are
\begin{eqnarray}
\Gamma^{(2)}(\vec{k},\omega_1,\omega_2)&=&\Gamma^{(0)}(\vec{k},\omega_1,\omega_2)-\nonumber\\
& &\displaystyle(n-2)\frac{1}{2}\lambda^2\sum_{\omega}\int
d\vec{p}G_{0}(\vec{p},\omega,\omega_1)G_{0}(\vec{k}-\vec{p},\omega_2,-\omega)\nonumber\\
\label{15}
\end{eqnarray}
where
\begin{eqnarray}
\Gamma^{(0)}(\vec{k},\omega_1,\omega_2)&=&
t+k^2+s^2(\omega_1^2+\omega_2^2)=\nonumber \\
& & =G^{-1}_{0}(\vec{k},\omega_1,\omega_2) \label{16}
\end{eqnarray}
and
\newpage
\begin{eqnarray}
\Gamma^{(3)}(\vec{k_1},\vec{k_2},\omega_1,\omega_2,\omega_3)=
\lambda +\nonumber \\
\displaystyle(n-3)\lambda^3\sum_{\omega}\int
d\vec{p}G_{0}(\vec{p},\omega_1,\omega)G_{0}(\vec{k_1}+\vec{p},-\omega,\omega_2)
G_{0}(\vec{k_1}+\vec{k_2}+\vec{p},-\omega,\omega_3)\nonumber\\
\label{17}
\end{eqnarray}
The theory will be renormalized at the critical point $t=0$. To
get away from the critical point we should consider a perturbation
expansion in $t$ by means of the insertion\cite{amit}
\begin{eqnarray}
\Delta A=\frac{1}{2!}\sum_{\gamma,\nu}\sum_{\omega_1 \omega_2}
\int d\vec{q}t(\vec{q})\int
d\vec{p}Q_{\gamma\nu}(\vec{p},\omega_1,
\omega_2)Q_{\nu\gamma}(\vec{q}-\vec{p},-\omega_2,-
\omega_1)\nonumber \\
\label{18}
\end{eqnarray}
that leads to a third singular vertex function $ \Gamma^{(2,1)} $
with two external legs and one insertion shown in fig.(2)(bottom).
\begin{eqnarray}
\Gamma^{(2,1)}(\vec{k},\vec{q},\omega_1,\omega_2)=
1 +\nonumber \\
\displaystyle(n-2)\lambda^2\sum_{\omega}\int
d\vec{p}G_{0}(\vec{p},\omega,\omega_1)G_{0}(\vec{q}-\vec{p},-\omega_1,-\omega)
G_{0}(\vec{k}+\vec{p},\omega,\omega_2)\nonumber\\
\label{19}
\end{eqnarray}
At finite temperature $T$ and critical $t=0$, the sums over
Matsubara frequencies have only one singular term with $\omega =
0$ and the vertex functions are singular when $\omega_i = 0$, then
we recover the transition for classical spin glasses described by
an expansion in $ \epsilon= 6-d$.\cite{harris}\\
A different scenario emerges when $ T $ is near zero. For
sufficiently low $T$ the frequency sums may be replaced by
integrals
\begin{eqnarray}
\sum_{\omega}\longrightarrow \beta\int_{-\infty}^{\infty}d\omega
\label{20}
\end{eqnarray}
and now all the frequencies contribute to the renormalization
process and the vertex functions in eq.(\ref{15}),eq.(\ref{17})and
eq.(\ref{19}) will be singular at a new effective dimension
$D_c=d_c+1 =6$, the new critical space dimensionality becoming
$d_c=5$, as predicted\cite{zazie,hertz,sachdev,belitz}.
\newpage
\section*{3.\ Results}
In the following we present results for the critical properties in
an expansion in $\epsilon^\prime = 5-d $,to one loop order. The
new features that emerge from the calculation are that the
frequencies renormalize differently than the momenta, then the
exponent $z$ differs from unity and from the exponent $\eta$,
depending also on the dimensionality through the $\epsilon^\prime$
expansion. The integrals over momentum and frequency of
eq.(\ref{15}),eq.(\ref{17})and eq.(\ref{19}) are calculated in the
Appendix at a space dimensionality $d$, when they
converge\cite{amit,theumanngusmao} and the singularities appear as
dimensional poles in $\epsilon^\prime$. We obtain for the singular
parts, to leading order in the coupling constant
\begin{eqnarray}
\Gamma^{(2)}_{\alpha \gamma}(\vec{k},\omega_1,\omega_2)&=&k^2 + s^2(\omega_1^2+\omega_2^2)+\nonumber\\
& &(n-2)\frac{1}{6s
\epsilon^\prime}u_0^2\bigl[k^2+3s^2(\omega_1^2+\omega_2^2)\bigr]= \nonumber \\
& &\bigl[1+(n-2)\frac{1}{6s\epsilon^\prime}u_0^2\bigr]\bigl\{k^2 +
s^2(\omega_1^2+\omega_2^2)\bigl[1+\frac{n-2}{3\epsilon^\prime}u_0^2\bigr]\bigr\}
\nonumber \\
 \label{21}
 \end{eqnarray}
 \vspace{0.3cm}
\begin{eqnarray}
\Gamma^{(3)}_{\alpha \gamma \delta}=
u_0\kappa^{\epsilon^\prime/2}\big[1+(n-3)u_0^2\frac{1}{s\epsilon^\prime}\big]\nonumber
\\
\label{22}
\end{eqnarray}
\vspace{0.3cm}
\begin{eqnarray}
\Gamma^{(2,1)}_{\alpha \gamma}=
1+(n-2)u_0^2\frac{1}{s\epsilon^\prime}\nonumber
\\
\label{23}
\end{eqnarray}
where we introduced the bare dimensionless coupling $u_0$ through
\begin{eqnarray}
\lambda^2 \beta S_{d+1} = u_0^2 \kappa^{\epsilon^\prime}
\label{24}
\end{eqnarray}
and $ S_{d+1}$ is the surface of the unit sphere in $ d+1$
dimensions.In $\Gamma^{(3)}$ and $\Gamma^{(2,1)}$ the external
momenta and frequencies were taken at the symmetry point $
k_1^2=k_2^2= -2\vec{k_1}.\vec{k_2} = \omega_i^2 =\kappa^2 $, where
$\kappa$ is the scale parameter\cite{amit}.In order to cancel the
dimensional poles we must introduce a renormalized, dimensionless,
coupling $u$ and renormalized vertex functions by means of
renormalization of the field $Q_{\alpha \gamma}$ and of the
insertion $Q_{\alpha \gamma}^2$ through the functions $Z_Q $ and
$Z_{Q^2} $. The correction to the frequency term $\omega_i^2 $ in
$\Gamma^{(2)}$ in eq.(\ref{21}) is different than the contribution
to $k^2$, then besides the field renormalization $Z_Q $ that keeps
the coefficient of $k^2$ equal to unity it is necessary to
renormalize also the frequency coefficient $ s(u) $.All together
we obtain
\begin{eqnarray}
\Gamma^{(2)}_R(u)=Z_Q(u)\Gamma^{(2)}(u_0,s) \nonumber \\
\Gamma^{(3)}_R(u)=Z_Q^{\frac{3}{2}}(u)\Gamma^{(3)}(u_0) \nonumber \\
\Gamma^{(2,1)}_R(u)=\overline{Z}_{Q^{2}}(u)\Gamma^{(2,1)}(u_0) \nonumber \\
\label{25}
\end{eqnarray}
where, in the interesting limit $n=0$
\begin{eqnarray}
u_0=u\big[1+\frac{5}{s\epsilon^\prime}u^2 \big]\hspace{0.5cm} (a) \nonumber \\
Z_Q(u)=1+\frac{1}{3\epsilon^\prime}u^2 \hspace{0.5cm} (b)\nonumber \\
\overline{Z}_{Q^{2}}(u)=1+ \frac{2}{\epsilon^\prime}u^2
\hspace{0.5cm} (c)\nonumber
\\
s^2 =1+\frac{2}{3\epsilon^\prime}u^2\hspace{0.5cm} (d) \nonumber
\\
\label{26}
\end{eqnarray}
From eq.(\ref{26}-a)we calculate the $\beta$-function
\begin{eqnarray}
\beta(u)&=& \kappa \frac{\partial u}{\partial \kappa}\Big|_\lambda
\nonumber \\
&= &-\frac{\epsilon^\prime}{2} u\big[1-\frac{5}{\epsilon^\prime}u^2
\big]\nonumber \\
\label{27}
\end{eqnarray}
that vanishes at the trivial fixed points $u^\ast = 0 $, stable for
$\epsilon^\prime < 0 $, and ${u^\ast}^2 = \frac{1}{5}\epsilon^\prime
$,  stable for $\epsilon^\prime > 0 $. To obtain the critical
exponents we have to solve the renormalization group
equations\cite{amit} for the vertex function
$\Gamma^{(2)}_R(\vec{k},s\omega_i,t,u, \kappa)$ near criticality,
where $t=\frac{\mu-\mu_c}{\mu_c} $. Now we have to take into account
also the dependence of $ s $ on $\kappa$ through the coupling $u$,
so calling $y_i = s\omega_i , i=1,2 $, we obtain the renormalization
group equation at the fixed point $ \beta(u^\ast) =0 $
\begin{eqnarray}
\Big[\kappa\frac{\partial}{\partial \kappa}+
\gamma_s^\ast\sum_iy_i\frac{\partial}{\partial y_i} - \theta
t\frac{\partial}{\partial t}-
\eta\Big]\Gamma^{(2)}_R(\vec{k},y_i,t,\kappa)= 0 \nonumber \\
\label{28}
\end{eqnarray}
where
\begin{eqnarray}
\eta =\Big[\kappa\frac{\partial}{\partial
\kappa}\ln{Z_Q}\Big]_{u=u^\ast}\nonumber \\
 \vspace{0.3cm}
 \theta=\Big[\kappa\frac{\partial}{\partial
\kappa}\ln{Z_{Q^2}}\Big]_{u=u^\ast} - \eta \nonumber \\
 \vspace{0.3cm}
 \gamma_s^\ast =\Big[\kappa\frac{\partial}{\partial
\kappa}\ln{s}\Big]_{u=u^\ast}\nonumber \\
\label{29}
\end{eqnarray}
The solution for $\Gamma_R^{(2)} $ has the scaling form
\begin{eqnarray}
\Gamma^{(2)}_R(\vec{k},y_i,t,\kappa)= \kappa^\eta
\Phi\big[\vec{k};y_it\kappa^{\theta -\gamma_s^\ast}\big] \nonumber
\\
\label{30}
\end{eqnarray}
where $\Phi $ is a function of the joint variable $
y_it\kappa^{\theta -\gamma_s^\ast} $ and dimensional analysis
tells us that, for $ \rho $ an inverse length:
\begin{eqnarray}
\Gamma^{(2)}_R(\vec{k},y_i,t,\kappa)&=&
\rho^2\Gamma^{(2)}_R(\frac{\vec{k}}{\rho},\frac{y_i}{\rho},\frac{t}{\rho^2},
\frac{\kappa}{\rho}) \nonumber \\
&=& \rho^2\big[\frac{\kappa}{\rho}\big]^\eta
\Phi\big[\frac{\vec{k}}{\rho};\frac{y_it\kappa^{\theta
-\gamma_s^\ast}}{\rho^{3+\theta-\gamma_s^\ast}}\big] \nonumber
\\
\label{31}
\end{eqnarray}
If we choose\cite{amit}
\begin{eqnarray}
\rho= \kappa\big(\frac{t}{\kappa^2}\big)^{\frac{1}{2+\theta}}
\label{32}
\end{eqnarray}
we obtain
\begin{eqnarray}
\Gamma^{(2)}_R(\vec{k},y_i,t,\kappa)=
 \kappa^2\big(\frac{t}{\kappa^2}\big)^{\nu(2-\eta)}
\Phi\big[\frac{\vec{k}}{\kappa}(\frac{t}{\kappa^2})^{-\nu};\frac{y_i}{\kappa}
(\frac{t}{\kappa^2})^{-\nu z}\big] \nonumber
\\
\label{33}
\end{eqnarray}
from where we identify the space correlation length exponent
\begin{eqnarray}
\xi= (\frac{t}{\kappa^2})^{-\nu} \hspace{0.5cm} \nu^{-1}= 2+
\theta\nonumber \\
\label{34}
\end{eqnarray}
and the time correlation length exponent
\begin{eqnarray}
\xi_t= (\frac{t}{\kappa^2})^{-\nu z}=\xi^z \hspace{0.5cm}
z=1-\gamma_s^\ast   \nonumber \\
\label{35}
\end{eqnarray}
From eq.(\ref{26}),eq.(\ref{29}),eq.(\ref{34}) and eq.(\ref{35})
we obtain the results for the critical exponents, at the
non-trivial fixed point
\begin{eqnarray}
\eta=-\frac{1}{15}\epsilon^\prime;\hspace{0.5cm}\nu=\frac{1}{2}+
\frac{1}{12}\epsilon^\prime;\hspace{0.5cm}z=1+\frac{1}{15}\epsilon^\prime
\nonumber \\
\label{36}
\end{eqnarray}
\newpage
\section*{4.\ Conclusions}
In the present paper we analyze the critical properties of a
quantum spherical spin glass model with short range, random
interactions. Since the model allows for rigorous detailed
calculations, we can show how the effective partition function
calculated with help of the replica method for the spin glass
fluctuating fields $ Q_{\alpha \gamma}(\vec{k}, \omega_1,
\omega_2) $ separates into a mean field contribution for the
$Q_{\alpha \alpha}(0, \omega,  -\omega) $ and a strictly short
range partition function for the fields $ Q_{\alpha \neq
\gamma}(\vec{k}, \omega_1, \omega_2) $.Here $ \alpha, \gamma =1..n
$ are replica indices. The mean field part $ W_{MF}$ coincides
with previous results\cite{pedroalba} and a saddle point
calculation provides the phase diagram in fig.(1), as it is
discussed in the Appendix. We stress that $Q_{\alpha \alpha}(0,
\omega, -\omega) $ is not an order parameter, as it does not
vanish above the transition temperature, and the order parameter
in the quantum spherical, infinite range, spin glass identically
vanish\cite{pedroalba}. The short range part $W_{SR}$ describes a
phase transition in a $ Q^3 $-field theory, where the fluctuating
fields depend on a space variable $\vec{r} $ and two times $
\tau_1$ and $\tau_2$. This we analyze using the renormalization
group with dimensional regularization and minimal subtraction of
dimensional poles\cite{amit}. By generalizing the method in
ref.(\cite{amit}) to our particular situation, we can identify the
critical dimensionality as $ d_c = 5$ at very low temperatures due
to the dimensionality shift $D_c = d_c+1 =6 $. We then perform an
$ \epsilon^\prime $ expansion to order one loop to calculate the
critical exponents by solving the renormalization group equations,
and they are listed in eq.(\ref{36}).\\
A general Landau theory of quantum spin glasses of M-components
rotors was presented in ref.(\cite{readsachdev}). Based on general
properties of symmetry and invariance, the authors present an
effective functional for spin glass $Q$-fields, and at some points
we make contact with their results. Our fields, as theirs, are
bilocal in time, but our result for the effective functional is
simpler and more tractable by standard field theory methods. It is
well known\cite{stanley} than the classical, non-random spherical
model is equivalent to the $M \rightarrow \infty $ limit of the
$M$-vector model. The same equivalence holds between the
infinite-range spherical spin glass and the infinite-range
$M$-vector spin glass in the classical\cite{koster,jairo} and in the
quantum case\cite{ye,pedroalba}. A particular feature of the
infinite-range spherical spin glass is that it can be solved exactly
without need of the replica method\cite{koster} because annealing is
exact in this model due to the internal Becchi-Rouet-Stora-Tyutin
(BRST) supersymmetry\cite{pedroalba} and as a consequence Ward
identities tell us that the order parameter identically vanish. In
the case of the short-range quantum spherical spin-glass considered
here, we showed that replicas are needed and that the partition
functional separates exactly into a mean-field part for the replica
diagonal $Q_{\alpha \alpha}(k=o,\omega,-\omega) $ and a short range
part for the fluctuating  $Q_{\alpha \neq\beta}(k,\omega_1,\omega_2)
$ in Eq.(\ref{11}), while in the spin glass of $M$-components
quantum rotors with short range disorder considered in
Ref.(\cite{readsachdev}) the replica diagonal $ Q_{\alpha
\alpha}(\omega) $ is considered as an order parameter and a Landau
functional is constructed for fluctuations diagonal in replica space
around it. This leads to a theory where the time derivatives and the
critical mass appear in the {\it linear}, in place of quadratic,
term in the effective action. As a consequence of having different
interactions, the renormalization group equations and critical
exponents turn out to be $M$- independent, and the critical
dimensionality obtained in Ref.(\cite{readsachdev}) also differs
from ours. We conclude this is due to the fact that, in the case of
short-range disorder considered here ,the quantum spherical
spin-glass model belongs to a different universality class than the
$M $-components quantum rotors model in Ref.(\cite{readsachdev}).\\
We may ask how our results would apply to the quantum $p$-spin
spherical model theories in Ref.(\cite{theo,serral}), in the case of
short range disorder. In these theories the action depends on the
first time derivative of the fields, then the inverse propagators
would have a linear dependence on frequency (and not quadratic as it
is the case here), so the results of a RG calculation remain open.
\section*{\ Acknowledgement}
We thank W.K.Theumann for discussions. We gratefully acknowledge
financial support from FAPERGS and CNPq.
\newpage
\section*{\ Appendix}
1.\underline{ Effective functional}\\
We derive here the functional $W_n$ in eq.(\ref{9}). We obtain by
replicating ${\cal W}$ in eq.(\ref{7}) and averaging over $
P(J_{ij}) $ in Eq.(\ref{3})
\begin{eqnarray}
 W_n= &&\displaystyle \int \prod_{i\alpha}{\cal D} S_{i \alpha}
\exp{\left(-\cal A{_{O} - A_{SG}} \right)} \label{37}
\end{eqnarray}
where $\alpha=1,2,...n $ is the replica index and the free action
$ \cal{A_0}$ is given by
\begin{eqnarray}
{ \cal A_{O}} =\int_{0}^{\beta} d \tau  \displaystyle
\sum_{i}\sum_{\alpha}\left(\frac{I}{2} \left(\frac{\partial S_{i
\alpha}}{\partial \tau}\right)^{2} + \mu S_{i
\alpha}^2(\tau)\right) \label{38}
\end{eqnarray}
while  the interacting part is
\begin{eqnarray}
{ \cal A_{SG}} =\frac{J^2}{4} \displaystyle\sum_{i,j}\sum_{\alpha
\alpha^\prime}V_{ij}\int_{0}^{\beta} d \tau \int_{0}^{\beta} d
\tau^\prime S_{i \alpha}(\tau)S_{j \alpha}(\tau)
S_{i\alpha^{\prime}}(\tau^{\prime})S_{j
\alpha^{\prime}}(\tau^{\prime})\nonumber \\ \label{39}
\end{eqnarray}
We introduce the spin glass fields $Q_{i \alpha \gamma}(\tau,
\tau^\prime) $ by splitting the quartic term by means of a
Stratonovich-Hubbard transformation and we obtain
\begin{eqnarray}
 W_n= \displaystyle \int\prod_{i} \prod_{\alpha \gamma}{\cal D}Q_{i \alpha
 \gamma}(\tau,
\tau^{\prime})\nonumber \\
\exp{\big[-\frac{J^{2}}{4}\sum_{\alpha
\gamma}\int_{0}^{\beta}\int_{0}^{\beta}
 d \tau d \tau^{\prime}\sum_{i,j}Q_{i \alpha \gamma}(\tau ,\tau^{\prime})V_{i,j}^{-1}
 Q_{j \gamma \alpha}(\tau ,\tau^{\prime})\big]}
 \exp{[N \Lambda]}\nonumber \\
 \label{40}
\end{eqnarray}
where
\begin{eqnarray}
\exp{[N\Lambda]}=\int\prod_{i, \alpha}{\cal D}S_{i
\alpha}(\tau)\nonumber \\
\exp{\left[-{\cal A_{O}}
-\frac{J^{2}}{2}\sum_{\alpha
\gamma}\int_{0}^{\beta}\int_{0}^{\beta}d \tau d \tau^{\prime}
\sum_{j}Q_{j \alpha \gamma}(\tau, \tau^{\prime})S_{j
\alpha}(\tau)S_{j \gamma}(\tau^{\prime})\right]} \nonumber \\
\label{41}
\end{eqnarray}
In eq.(\ref{41}) appear the fields $ Q_{j \alpha \gamma}(\tau,
\tau^{\prime}) $ depending on two independent times $ \tau,
\tau^\prime $    and  \underline {not} on the time difference. We
define the space and time Fourier transform
\begin{eqnarray}
S_{\alpha}(\vec{k} \omega)=\frac{1}{\beta \sqrt{N}}\displaystyle
\int_0^\beta d\tau \sum_jS_{j
\alpha}(\tau)\exp{-i[\vec{k}.\vec{R}_j+\omega \tau]}\nonumber \\
\label{42}
\end{eqnarray}
\begin{eqnarray}
Q_{\alpha \gamma}(\vec{k} \omega \omega^\prime)=\frac{1}{\beta^2
 N}\displaystyle \int_0^\beta d\tau \int_0^\beta d\tau^\prime
\sum_j Q_{j\alpha \gamma}(\tau
\tau^\prime)\exp{-i[\vec{k}.\vec{R}_j+\omega \tau
+\omega^\prime \tau^\prime]}\nonumber \\
\label{43}
\end{eqnarray}
where $\omega= \frac{2\pi m}{\beta} $ are bosonic Matsubara
frequencies and we obtain from eq.(\ref{40})-eq.(\ref{43})
\begin{eqnarray}
 W_n= \displaystyle \int\prod_{\vec{k}\omega_1 \omega_2} \prod_{\alpha \gamma}
 dQ_{ \alpha
 \gamma}(\vec{k} \omega_1 \omega_2)\nonumber \\
\exp{\Big[-\frac{(\beta J)^2}{4}\sum_{\alpha
\gamma}\sum_{\vec{k}\omega_1 \omega_2}Q_{ \alpha \gamma}(\vec{k}
\omega_1 \omega_2)V_(\vec{k})^{-1}Q_{ \alpha \gamma}(-\vec{k},
-\omega_1,- \omega_2)\Big]}
\exp{[N \Lambda]}\nonumber \\
 \label{44}
\end{eqnarray}
where
\begin{eqnarray}
 \exp{[N\Lambda]}=\displaystyle \int\prod_{\alpha}
\prod_{\vec{k}\omega}dS_{\alpha}(\vec{k} \omega)
\exp{\Big[-\sum_{\alpha}\sum_{\vec{k}\omega}\big(\frac{\beta
I\omega^2}{2} + \mu \beta \big) S_{\alpha}(\vec{k}
\omega)S_{\alpha}(-\vec{k} -\omega)\Big]} \nonumber \\
\exp{\Big[\frac{(\beta J)^{2}}{2}\sum_{\alpha \gamma}
\sum_{\vec{k}\vec{q}}\sum_{\omega \omega^\prime}Q_{\alpha
\gamma}(\vec{q}\omega \omega^\prime)S_{\alpha}(\vec{k}\omega)
S_{\gamma}(\vec{k}-
\vec{q}, \omega^\prime)\Big]}  \nonumber \\
\label{45}
\end{eqnarray}
and for short range forces $ V(k)^1 = 1+k^2 $. The next step is to
separate the term with $Q_{\alpha \alpha}(0,\omega,-\omega)$ in
eq.(\ref{45}) that can be introduced into the free action for $
S_{\alpha}(\vec{k} \omega)$, with the result
\begin{eqnarray}
W_n = W_{MF} W_{SR} \label{46}
\end{eqnarray}
where $ W_{MF}$ is the mean field partition functional for the $
Q_{\alpha}(\omega)=Q_{\alpha \alpha}(0\omega -\omega) $ mode
\begin{eqnarray}
 W_{MF}= \displaystyle \int\prod_{\omega \alpha}
 dQ_{ \alpha}(\omega)
\exp{-\frac{N}{2}\sum_{\alpha \omega}\Big[\frac{(\beta
J)^2}{2}Q_{\alpha}^2( \omega)+ \ln{\big(\beta I \omega^2 + 2\beta
\mu -(\beta J )^{2} Q_{ \alpha}(\omega)\big)}\Big]} \nonumber \\
\label{47}
\end{eqnarray}
and $W_{SR} $ is the partition functional for the fluctuations
$Q_{\alpha \neq \gamma}$
\begin{eqnarray}
W_{SR} = \displaystyle\int\prod_{\alpha \neq \gamma}\prod_{\vec{k}
\omega} dQ_{\alpha \neq \gamma}(\vec{k}
\omega_1 \omega_2)\exp{-[{\cal A}_{free} +{\cal A}_{int}]} \nonumber \\
\label{48}
\end{eqnarray}
\begin{eqnarray}
{\cal A}_{free}=N \sum_{\vec{k} \omega_1 \omega_2}\sum_{\alpha
\gamma}Q_{\alpha \gamma}(\vec{k} \omega_1 \omega_2)Q_{\gamma
\alpha}(-\vec{k}, -\omega_1, -\omega_2)(\beta J
)^2 \Gamma_0(\vec{k} \omega_1 \omega_2)\nonumber \\
\label{49}
\end{eqnarray}
\begin{eqnarray}
{\cal A}_{int}= \frac{(\beta J )^6}{3!}N
\sum_{\vec{k}_1\vec{k}_2}\sum_{\alpha \omega_1}\sum_{\gamma
\omega_2}\sum_{\delta \omega_3}\nonumber \\
Q_{\alpha
\gamma}(\vec{k}_1 \omega_1 \omega_2)Q_{\gamma \delta}(\vec{k}_2,
-\omega_2, \omega_3)Q_{\delta \alpha}(-\vec{k}_1 -\vec{k}_2,
-\omega_3,
-\omega_1)\prod_{i}g_0(\omega_i) \nonumber \\
\label{50}
\end{eqnarray}
where
\begin{eqnarray}
\Gamma_0(\vec{k} \omega_1 \omega_2)= 1+q^2-(\beta
J)^2g_0(\omega_1)g_0(\omega_2) \label{51}
\end{eqnarray}
The function $g_0(\omega)$ in eq.(\ref{50}) is the momentum
independent, non-interacting two point function for the field
$S_{\alpha}(\omega)$
\begin{eqnarray}
g_0(\omega)=\frac{2}{\beta I \omega^2 + 2\beta \mu -(\beta J )^{2}
Q_{ \alpha}(\omega)} \label{52}
\end{eqnarray}
The variables $Q_{\alpha \alpha}(\vec{q} \neq 0, \omega_1 \omega_2)
$ are not critical and are not coupled to the spin glass field, so
we ignore them.
\\
2.\underline{ Mean Field Solution} At the saddle point of $W_{MF}
$ in eq.(\ref{47}) we obtain
\begin{equation}
2Q_\alpha(\omega) = g_0(\omega) \label{53}
\end{equation}
The mean spherical condition of eq.(\ref{5}) reduces to
\begin{eqnarray}
-\frac{1}{n}\frac{\partial}{\partial\mu}\ln{W_{MF}} = \beta N
\nonumber
\end{eqnarray}
and it gives at the saddle point
\begin{eqnarray}
\displaystyle
\int_{L_{-}}^{L_{+}}dy\sqrt{(L_{+}^2-y^2)(y^2-L_{-}^2)}\coth(\frac{\beta
y}{2\sqrt{I}})= 2\pi J^{2}\sqrt{I} \label{54}
\end{eqnarray}
where
\begin{eqnarray}
L_{\pm}^2 = 2\mu \pm 2J \label{55}
\end{eqnarray}
that is just the condition found previously by us \cite{pedroalba}
for the mean field quantum spin glass and it gives $ \mu$ as a
function of $ T $ and $ I $. For high temperatures, the chemical
potential $ \mu  \rightarrow \infty $, while $ \mu = \mu_c = J $
at the critical temperature $ T_c(I) $ and the critical value
$I_c$ is reached when $ T_c(I_c) = 0 $, as it is shown in the
phase diagram in fig.(1). The high $\mu$ (high temperature )
solution for $ Q_\alpha(\omega) $ in eq.(\ref{53}) gives for
$\Gamma_0(\vec{k} \omega_1 \omega_2)$ in eq.(\ref{51}), when $ \mu
> J $
\begin{eqnarray}
\Gamma_0(\vec{k} \omega_1 \omega_2)=  1-(J/\mu)^2
+\frac{I}{2J}(\omega_1^2 +\omega_2^2) + q^2 \label{56}
\end{eqnarray}
Introducing eq.(\ref{56}) in eq.(\ref{50}), rescaling the fields
$Q_{\alpha \gamma} \rightarrow \frac{1}{\beta JN}Q_{\alpha
\gamma}$ and using $g_0(\omega=0,\mu=\mu_c) = (\beta J)^{-1} $, we
arrive to the effective spin glass partition functional in the
main text. We took explicitly the continuum limit in real space by
replacing, for vanishing lattice constants
\begin{eqnarray}
\frac{1}{N}\sum_{\vec{k}} \rightarrow \int d\vec{k} \nonumber \\
\end{eqnarray}
while for finite temperature the sum over Matsubara frequencies $
\omega =\frac{2\pi m}{\beta}$ are over the discrete index $m$. We
discuss next the regions with $ T> T_c $ and $I < I_c $.\\
a. {\it Classic Paramagnet (high temperature) }:
$\frac{\beta}{\sqrt{I}}\rightarrow 0$\\
 In this limit we are in the classical region and the integral in
 eq.(\ref{54}) can be solved exactly\cite{pedroalba} with the
 result
 \begin{equation}
 2(\frac{\mu}{J} -1)= (\frac{1}{\sqrt{\beta J}}-\sqrt{\beta
 J})^2 \label{58}
 \end{equation}
b. {\it Quantum Paramagnet (low temperature) }:
$\frac{\beta}{\sqrt{I}}\rightarrow \infty $ \\
In this limit $\coth{\frac{\beta y}{2\sqrt{I}}} \approx 1 $ and
the integral in eq.(\ref{54}) can be solved in terms of elliptic
integrals. For $(\frac{\mu}{J} -1) << 1$ we obtain
\begin{eqnarray}
2\pi[\sqrt{I_c J}-\sqrt{I J}]\approx -\frac{4}{3}(\frac{\mu}{J}
-1)\ln{[\frac{\mu}{J} -1]} \label{59}
\end{eqnarray}
Introducing eq.(\ref{58}) into eq.(\ref{59})we obtain the dotted
curve in fig.(1) that separates the classical from quantum
paramagnetic regions.\\
\\
3.\underline{ Integrals in dimensional regularization} In the low
temperature limit the sum over frequencies are replaced by
integrals as indicated in eq.(\ref{20}), then we need for
$\Gamma^{(2)}$ in eq.(\ref{15}), at the critical value $ t=0
$\cite{amit}
\begin{eqnarray}
I_2&=& \displaystyle \int d\omega d\vec{p}\frac{1}{p^2
+s^2(\omega_1^2+\omega^2)} \frac{1}{[\vec{p}-\vec{k}]^2
+s^2(\omega^2+\omega_2^2)} \nonumber \\
&=& \frac{S_{d+1}}{2s}\Gamma(\frac{d+1}{2})\Gamma(\frac{3-d}{2})
\int_0^1dx\frac{1}{[x(1-x)k^2+xs^2\omega_1^2+(1-x)s^2\omega_2^2]^{\frac{3-d}{2}}}
\nonumber \\ \label{60}
\end{eqnarray}
where $S_{d+1}$ is the surface of the unit sphere in
$(d+1)$-dimensions. we can see that $ \Gamma(\frac{3-d}{2})$ has a
dimensional pole at $ d_c = 5$, then calling $ \epsilon^\prime =
5-d$ we obtain the singular contribution in eq.(\ref{21}).\\
To renormalize $\Gamma^{(3)}$ and $\Gamma^{(2,1)}$ in
eq.(\ref{17})and eq.(\ref{19})we need to calculate
\begin{eqnarray}
I_3= \displaystyle \int d\omega d\vec{p} \nonumber \\
\frac{1}{[p^2+s^2(\omega_1^2+\omega^2)]}
\frac{1}{[(\vec{p}+\vec{k_1})^2
+s^2(\omega^2+\omega_2^2)]}\frac{1}{[(\vec{p}+\vec{k_1}+\vec{k_2})^2
+s^2(\omega^2+\omega_3^2)]} \nonumber \\
\label{61}
\end{eqnarray}
what we do by taking the external momenta and frequencies at the
symmetry point\cite{amit}
\begin{eqnarray}
k_1^2=k_2^2=\omega_i^2 = \kappa^2;
\hspace{0.6cm}\vec{k_1}.\vec{k_2}=-\frac{\kappa^2}{2}\nonumber \\
\label{62}
\end{eqnarray}
and performing the integral in $d+1$-dimensions as in
eq.(\ref{60}) with the result
\begin{eqnarray}
I_3&=&
\frac{S_{d+1}}{s}\frac{\Gamma(\frac{d+1}{2})\Gamma(\frac{5-d}{2})}{\Gamma(3)}
\kappa^{-(5-d)}\times \nonumber \\
&&\int_0^1dx_1\int_0^{1-x_1}dx_2\frac{1}{[x_1(1-x_1)+x_2(1-x_1-x_2)
+s^2]^{\frac{5-d}{2}}} \nonumber \\
\label{63}
\end{eqnarray}
We see again the dimensional pole in $\Gamma(\frac{5-d}{2})$ at
$d_c= 5 $, then considering the singular part at the pole in
$\epsilon^\prime = 5-d $, we obtain the results in eq.(\ref{22})
and eq.(\ref{23}).
\newpage
\section*{5.\ Figure Captions}
Fig.1 Phase diagram in the $T vs 1/I $ plane. a) Critical line $
T_c(1/I) $( full) separating the classical paramagnetic(top) from
the spin glass phase(bottom).
 b)Estimated line (dots) separating the classical paramagnetic(top) from the
 quantum paramagnetic (bottom) regions.\\
\\
Fig.2 Diagrammatic representation of the vertex functions. A double
line represents a propagator with two replica indices $ \alpha,
\gamma $, momentum $\vec{k} $ and two frequencies $ \omega_1,
\omega_2 $. (a) Top: $\Gamma^{(2)}$ ; (b)Middle: $\Gamma^{(3)} $;
(c) Bottom: $\Gamma^{(2,1)}$.

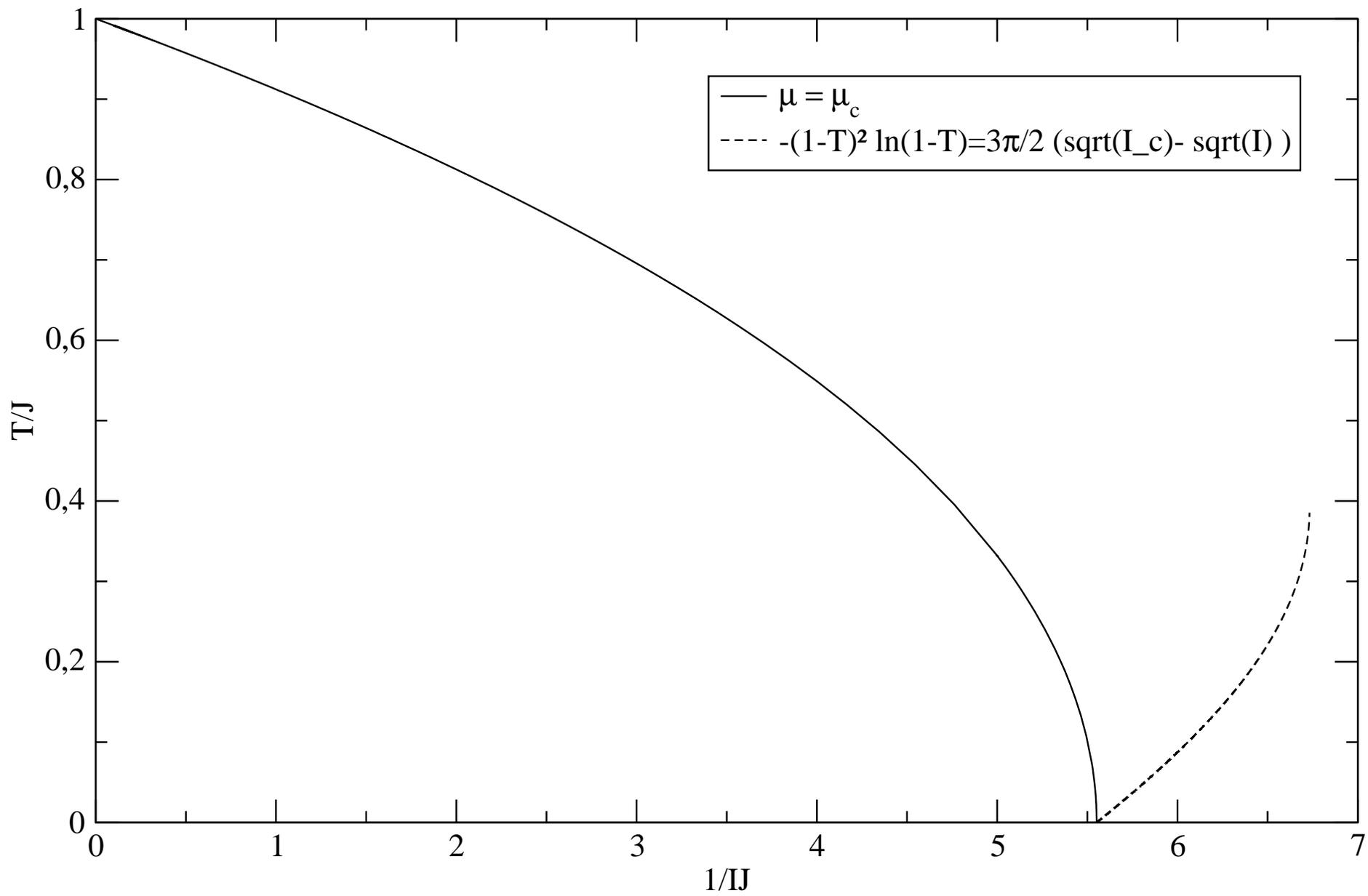
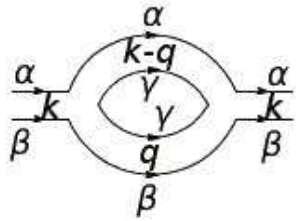
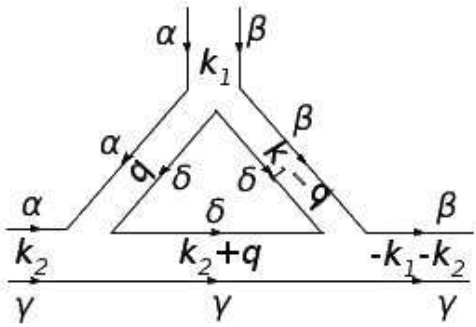
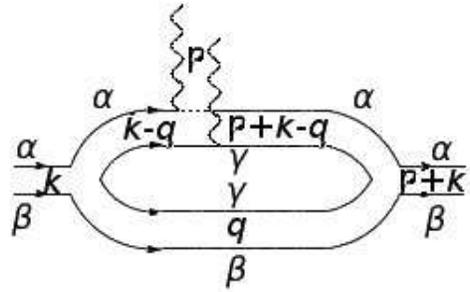
\end{document}